\documentclass[aps,amsmath,amssymb,twocolumn,10pt,superscriptaddress]{revtex4-1}
\usepackage[T1]{fontenc}
\usepackage{graphicx}  
\usepackage[dvipsnames]{xcolor}
\usepackage[colorlinks=true,linkcolor=blue,citecolor=red,urlcolor=magenta]{hyperref}
\usepackage{utopia} 
 
\usepackage{bm}

\begin{document} 

\title{Stability conditions for bound states in antiprotonic atoms}

\author{Fredrik Parnefjord Gustafsson}
\affiliation{EP Department, CERN, 1211 Geneva 23, Switzerland}
\author{Daniel P\k{e}cak}
\affiliation{Institute of Physics, Polish Academy of Sciences, Aleja Lotnikow 32/46, PL-02668 Warsaw, Poland}
\author{Tomasz Sowi\'nski}
\affiliation{Institute of Physics, Polish Academy of Sciences, Aleja Lotnikow 32/46, PL-02668 Warsaw, Poland}

\begin{abstract}
We study the stability of bound antiprotons in close proximity to the atomic nucleus. Using experimental data from X-ray spectroscopy measurements of antiprotonic atom transitions, we tune a minimal theoretical framework and estimate parameter ranges where the strong or electromagnetic decay channels govern the stability of the deepest bound states. In this way, we present an overview of the dominant decay mechanism of antiprotonic orbits for a given principle quantum number, to assist future spectroscopic experiments on antiprotonic atoms.
\end{abstract}

\maketitle
\section{Introduction}

Precision spectroscopy measurements of bound-state systems provide rigorous tests of fundamental forces and offer unique opportunities to search for effects that cannot be described by current theories. Highly charged ions, molecules, and exotic atoms, where bound particles are subject to extreme fields, serve as laboratories for probing these interactions with unprecedented accuracy \cite{safronova2018search, kozlov2018highly, paul2021testing, PhysRevLett.128.011802,PhysRevLett.128.112503}.
One particularly intriguing system for testing fundamental forces is the antiprotonic atom. In this system, the antiproton, with approximately 1836 times the mass of an electron, can form a bound state with a significantly reduced orbital radius near the atomic nucleus. These close orbits around the nucleus open a unique window into both strong-field quantum electrodynamics (QED) \cite{paul2021testing}, the short-ranged strong force interactions \cite{PhysRevLett.29.1132}, and may reveal phenomena beyond the Standard Model \cite{liu2025probing}. When the antiproton is captured by an atom it predominantly occupies high angular momentum Rydberg states, which limits its overlap with the nucleus due to the centrifugal barrier~\cite{Beck1993, Hori2011}. In these Rydberg states, the strong electromagnetic field gradient can be exploited to test QED contributions~\cite{paul2021testing,baptista2025towards}. As the antiproton cascades to lower energy levels, its wavefunction increasingly overlaps with the nucleus, significantly raising the probability of its inevitable annihilation. This drastically shortens the lifetime of the deeply bound states, leading to observable spectral line broadening, a phenomenon which can reveal details of the nuclear surface diffuseness \cite{PhysRevC.54.1832,PhysRevLett.87.082501}. However, the strong absorption itself introduces complexities. It can effectively repel the antiproton from the nuclear interior, leading to a non-linear 'saturation' effects. Consequently, the level width does not increase proportionally with the absorption strength. It is particularly true for the deepest bound states~\cite{NPAGalFriedmanSaturation, PLBGalFriedmanBatty, NPABattyFriedmanGalUnified}.

X-ray spectroscopy of antiprotonic atoms was first performed in the late 60's~\cite{BAMBERGER1970233} with further advanced studies performed in the LEAR facility at CERN from the 80's in to the 90's, yielding systematic data on the electromagnetic and strong interactions in these systems~\cite{DOSER2022103964,Batty1990,Backenstoss}. These experiments pioneered the use of X-ray spectroscopy as a unique method to investigate nuclear surface distributions across the nuclear chart~\cite{PhysRevC.54.1832,PhysRevLett.87.082501,Batty1990,Trzcinska2004}. Recent advancements, following the construction of the AD and ELENA antiproton decelerators at CERN, have enabled novel experimental techniques for the study of antiproton bound systems, such as experiments at
 ASACUSA probing antiprotonic helium using laser spectroscopy ~\cite{Hori2011,soter2022high}, pAX aiming at performing precision X-ray spectroscopy to test strong-field QED \cite{paul2021testing,baptista2025towards}, PUMA which focuses on the formation of antiprotonic atoms containing short-lived nuclei for the study of neutron-skin ~\cite{aumann2022puma}, and recently AEgIS, aiming at the laser triggered synthesis of antiprotonic atoms and the capture of annihilation fragment in a trapping environment~\cite{caravita2025long}.

In this work, we explore a global mapping of strong and electromagnetic contributions to the bound states of antiprotonic atoms using a minimal framework model. Acknowledging that precise quantitative predictions require more sophisticated models, we aim to provide a lowest-order, phenomenological overview. By analyzing existing data from LEAR-era antiprotonic atom X-ray spectroscopy, we estimate the strong-interaction decay rates for different principal quantum numbers within this minimal framework. We assume nuclear densities are described by the two-parameter Fermi distribution with an average nuclear diffuseness. The strong-decay rates, proportional to the nuclear overlap in our model, enable us to construct a chart highlighting the relative contributions of electromagnetic and strong interactions as functions of nuclear mass and antiproton principle quantum number. This chart identifies a transitional region where the stability is governed by the strong or electromagnetic interactions, highlighting different physics regions for future spectroscopic studies at a quantitative level.

\section{The model}
In our approach, we assume the simplest model of interplay between electromagnetic and strong interactions which allows us to draw some conclusions on the stability of the bound states of antiprotonic atoms. This model is built on four pillars describing {\bf (A)} the bound states of the orbiting antiproton, {\bf (B)} nuclear matter distribution, {\bf (C)} stability against electromagnetic decays, {\bf (D)} stability against strong interactions annihilation decays. Of course, each of these elements can be independently improved with more sophisticated approaches to get more accurate predictions. However for phenomenological reasoning, it is sufficient to make the following assumptions.

\subsection{Electromagnetic bound states of the antiproton}

We model the antiprotonic atom simply as a hydrogen-like system, where the antiproton of mass $m_{\bar{p}}$ orbits much heavier nucleus with charge $Z$. In the infinite nuclear mass approximation, the non-relativistic Hamiltonian describing orbiting antiproton has a form
\begin{equation} \label{Hamiltonian}
H = -\frac{\hbar^2}{2m_{\bar{p}}} \nabla^2 - \frac{Z e^2}{4\pi\epsilon_0}\frac{1}{r}.
\end{equation}
Eigenstates of this Hamiltonian are enumerated by quantum numbers $(n,\ell,m)$ and in spherical coordinates have a form
\begin{equation} \label{HydroWF}
    \psi_{n\ell m}(\bm{r}) = R_{n\ell}(r) Y_{\ell m}(\theta, \phi).
\end{equation}
Here, $Y_{lm}(\theta, \phi)$ are standard spherical harmonics while the radial component is expressed in terms of generalized Laguerre polynomials $\mathrm{L}_n^{(\alpha)}$ as

\begin{equation} \label{radialpart}
    R_{nl}(r) = {\cal N}_{n\ell} \left(\frac{2Zr}{a_0 n}\right)^\ell \, \mathrm{e}^{-Zr/a_0n}\,\mathrm{L}_{n-l-1}^{(2l+1)}\left(\frac{2Zr}{a_0 n}\right), 
\end{equation}

where the normalization factor is,
\begin{equation}
{\cal N}_{n\ell}=\frac{2}{n^2}\sqrt{\frac{Z^3(n-\ell-1)!}{a_0^3(n+\ell)!}}.
\end{equation}
Natural units of length and energy in this problem are provided by the Bohr length $a_0=\hbar/(\alpha m_{\bar{p}} c)\approx 28.8\,\mathrm{fm}$ and Bohr energy $E_0 = \alpha^2 m_{\bar{p}} c^2\approx 50\,\mathrm{keV}$, where dimensionless fine structure constant $\alpha=e^2/(4\pi\epsilon_0 \hbar c)\approx (137)^{-1}$. In this approximation eigenergies of the Hamiltonian \eqref{Hamiltonian} depend only on the principal quantum number $n$ and can be written as $E_n=-Z^2 E_0/(2 n^2)$. If we assume that the antiproton orbits the nucleus on quasi-classical circular Bohr orbit ($\ell\approx n-1$) then its spatial distribution is well-localized on a ring of the radius
\begin{equation}
r_n = \frac{a_0}{Z} n^2 = \frac{\hbar}{m_{\bar{p}}c}\frac{1}{Z\alpha}n^2.
\end{equation}
These states are preferred during the antiproton capture process, as has been shown experimentally \cite{Hori2011}. Therefore we restrict our focus to circular orbits in the following discussion.Of course, using pure hydrogen-like wavefunctions, is the zero order approximation which could be revised for more accurate predictions if needed. This is especially important for states having a large overlap with the nucleus, since then the antiproton's wavefunctions may be significantly distorted.

\vspace{0.5cm}
\subsection{Nuclear matter distribution}

We model the nuclear matter distribution of the nucleus with a spherically symmetric, equal for protons and neutrons, two-parameter Fermi distribution along the radial direction~\cite{schmidt1998nucleon}. In this case, the size of the nucleus is determined by the total number of nucleons $A$ and its characteristic radius is $R(A) = R_0 A^{1/3}$, where $R_0 = 1.2\,\mathrm{fm}$. The second parameter, the diffuseness of a nucleus surface $\sigma$, is of the order of  $R_0$ but its exact value will be determined later to tailor the model to the experimental data. With these two parameters the distribution of matter inside the nucleus is expressed as
\begin{equation} \label{nucleardensity}
\rho(A,r) = \frac{\rho_0}{1+\mathrm{exp}\left(\frac{r-R(A)}{\sigma}\right)},
\end{equation}
The amplitude $\rho_0$ is not a free parameter since it is always determined by normalization of the distribution to the total number of nucleons $A$. To set a direct connection between mass number $A$ with the charge of the nucleus $Z$ we approximate the neutron-to-proton ratio as $N/Z \approx 1 + \lambda A^{2/3}$, with $\lambda\approx 0.015$ being a ratio of droplet model coefficients related the Coulomb repulsion energy and the asymmetry energy~\cite{Royer2009}. This relations simply reflects the increasing neutron surplus required for nuclear stability as the size of the nucleus grows, due to the repulsive electrostatic forces between protons. From this relationship, the atomic number $Z$ can be empirically estimated from the liquid drop model charge $Z_{\mathrm{DM}}$ as
$Z \approx Z_{\mathrm{DM}} = A\left(2 + \lambda A^{2/3}\right)^{-1}$. We consequently use this approximation in the following.

\begin{figure*}
\includegraphics[width=0.9\linewidth]{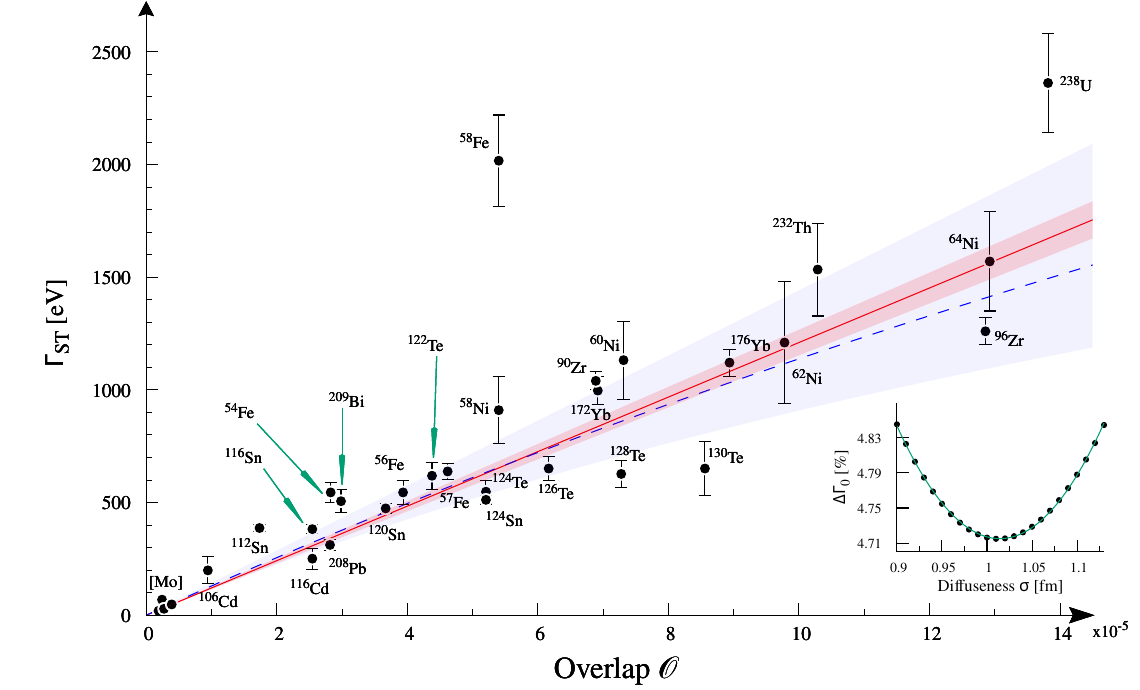}
    \caption{Strong interaction decay rate $\Gamma_\mathrm{ST}$ as a function of the antiproton-nucleus overlap ${\cal O}(n,A)$ (Eq. \ref{overlap}). The experimental decay rates acquired from \cite{kanert1986first, trzcinska2001information}. [Mo] represents isotopes of Molybdenum with the following mass numbers: 92, 94, 95, and 98. These data are fitted with a linear trend \eqref{overlap_approx} for optimized diffuseness $\sigma$ minimizing the uncertainty of obtained $\Gamma_0$ (as shown in the inset). The fit resulting from this optimization gives a value of $\hbar\Gamma_{0}=(12.1 \pm 0.6)\,\mathrm{MeV}$ and $\sigma=1.02\,\mathrm{fm}$, shown as a solid red line with the uncertainty represented by the shaded area. The blue dashed line and corresponding shaded area represent the fitting of an extended model \eqref{eq:gamma_sat} effectively taking into account saturation strong interaction saturation. See the main text and the Appendix for details.}
    \label{Fig1}
\end{figure*}

\subsection{Electromagnetic decays}

To estimate the effect of transitions induced by electromagnetic interactions on the stability of antiprotonic orbit, we only consider the dominant contribution to them. Due to the assumption of a circular Bohr orbit ($\ell\approx n-1$), the largest contribution comes from the electric dipole ($E1$) transitions between nearest orbits ($n \rightarrow n-1$). According to the Einstein formula a corresponding decay rate can be calculated as:
\begin{align}
    \Gamma_{\text{EM}}(n) &= \frac{\omega^3 |d|^2}{3\pi\epsilon_0 \hbar c^3} \nonumber \\
    &= \frac{Z^4\alpha^5}{6}\frac{m_{\bar{p}}c^2}{\hbar}\left[\frac{1}{(n-1)^2}-\frac{1}{n^2}\right]^3 {\cal D}^2(n),
\end{align}
where $n$-dependent numerical factor ${\cal D}^2(n)$ can be directly extracted from the dipole transition matrix element $|d|^2=e^2|\langle n,n-1||\boldsymbol{r}||n-1,n-2\rangle|^2=e^2 a_0^2Z^{-2}{\cal D}^2(n)$ and has a form
\begin{equation}
{\cal D}^2(n) = 2^{4n+1}n^{2n+2}\frac{(n-1)^{2n+4}}{(2n-1)^{4n+2}}.
\end{equation}
The lifetime estimated considering only this decay channel represents an upper limit to the real state lifetime. This calculation can be easily extended to include other decay channels.

\subsection{Decays induced by strong interactions}

The strong interaction is short-ranged, with an estimated range on the order of $1\,\text{fm}$~\cite{kanert1986first}. The strong interaction width $\Gamma_\mathrm{ST}$ is typically determined by the overlap of the antiproton's wavefunction $\psi(\bm{r})$ with the imaginary part of the strong interaction optical potential $V_\mathrm{opt}(\bm{r})$ which accounts for absorption (annihilation) \cite{BATTY1997385}:
\begin{equation} \label{true_gamma_st}
\Gamma_\mathrm{ST} = -2 \frac{\int\!\mathrm{d}^3r\,\psi^*(\bm{r}) \Im\!\left[V_\mathrm{opt}(\bm{r})\right] \psi(\bm{r})}{\int\!\mathrm{d}^3r\, |\psi(\bm{r})|^2}.
\end{equation}
In our minimal framework, we approximate this relationship using the assumption that $\Im\!\left[V_\mathrm{opt}(r)\right]$ is proportional to the nuclear density $\rho(A,r)$, modeled with a zero-range interaction \cite{Yukawa1935}. Moreover, we approximate the wavefunction $\psi(\bm{r})$ with the hydrogen-like wavefunction \eqref{HydroWF} which ignores the suppression effects caused by $\Im\!\left[V_\mathrm{opt}(\bm{r})\right]$. When the circular Bohr orbit is assumed ($\ell\approx n - 1$),  this leads to the expression where the annihilation rate is directly proportional to the overlap integral ${\cal O}(n,A)$ between the unperturbed antiproton density distribution and the nuclear density \eqref{nucleardensity}:
\begin{equation}\label{overlap_approx}
    \Gamma_\mathrm{ST}(n,A) \approx \Gamma_0\cdot {\cal O}(n,A).
\end{equation}
The overlap integral is given by:
\begin{equation}\label{overlap}
    {\cal O}(n,A) = \int_0^{\infty}\mathrm{d}r\, r^2 |R_{n,n-1}(r)|^2 \rho(A,r).
\end{equation}
The proportionality constant $\Gamma_0$ has to be extracted from experimentally estimated decay rates. In our work we perform this estimation based on measurements published in~ \cite{kanert1986first, trzcinska2001information}. In these works, strong interaction decay rates $\Gamma_\mathrm{ST}$ were extracted through several complementary methods, depending on the specific energy levels. For the lowest-lying observed transition, {\it i.e,} where the strong interaction effect is most pronounced, $\Gamma_\mathrm{ST}$ is extracted directly from the Lorentzian broadening of the X-ray lines, after accounting for instrumental broadening via a convolution with a Gaussian profile. For higher-lying levels, where direct broadening measurements are less precise, $\Gamma_\mathrm{ST}$ is deduced through an intensity balance method. This method compares the intensities of transitions populating and depopulating a given level, attributing any difference to the strong-interaction absorption.

\begin{figure*}
    \includegraphics[width=\textwidth]{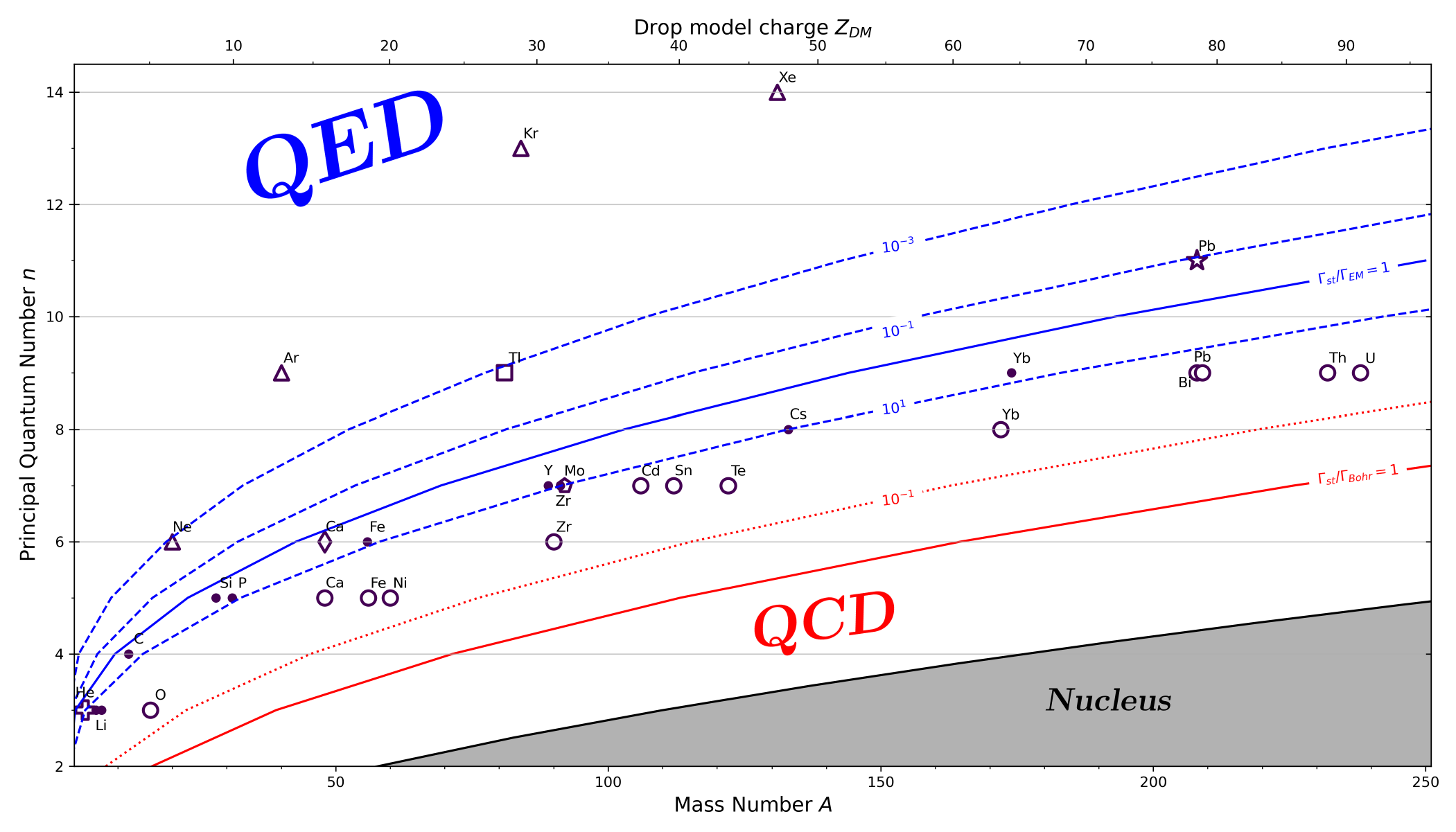}
    \caption{Experimentally observed x-ray transitions to the deepest observed principle quantum numbers $n$ in antiprotonic atoms of different nuclei. For clarity, we show only the lowest reported principal quantum numbers in different experiments (squares~\cite{bamberger1970observation}, crosses~\cite{poth1978antiprotonic}, pentagons~\cite{kanert1986first}, stars~\cite{kreissl1988remeasurement}, circles~\cite{trzcinska2001information},diamonds~\cite{hartmann2001nucleon}, triangles~\cite{gotta2008x}, points~\cite{PhysRevC.16.1945}). Red and blue isolines represent values of ratios $\Gamma_{\mathrm{ST}} / \Gamma_{\mathrm{Bohr}}$ and $\Gamma_{\mathrm{ST}} / \Gamma_{\mathrm{EM}}$ calculated within our minimal model, respectively. The QED and QCD labels mark regions of parameters where purely electromagnetic or strong effects appear dominant \emph{in this framework}. See the main text for further details.
    }
    \label{Fig2}
\end{figure*}

For each experimentally measured case, we determine the overlap ${\cal O}(n,A)$ according to \eqref{overlap}, with the assumption that the proportionality constant $\Gamma_0$ is measurement independent. We then perform a linear fit of the measured $\Gamma_\mathrm{ST}$ with respect to the numerical overlap ${\cal O}$ (see Fig.~\ref{Fig1}). Of course, the quality of this fit also significantly depends on the assumed nuclear diffuseness $\sigma$. Therefore, we additionally minimize the uncertainty of the determined $\Gamma_0$ by varying $\sigma$ and select the value that gives the minimal uncertainty (inset in Fig.~\ref{Fig1}).After performing this empirical procedure we find that this simple proportionality relation between decay rate $\Gamma_\mathrm{ST}$ and the overlap ${\cal O}$ holds very well on average for all elements considered, provided that one uses $\hbar\Gamma_0 = (12.1\pm 0.6)\,\mathrm{MeV}$ and $\sigma=1.02\,\mathrm{fm}$ for proportionality constant and diffuseness, respectively. In this sense, the minimal framework proposed is in good agreement with experimental data.

At this point, it should be mentioned, however, that we note some expected discrepancies from the simple linear trend visible in Fig.~\ref{Fig1}. They can be attributed to different approximations in our simplified model, originating in, {\it a.o.}, description of the nuclear density, using pure hydrogen-like wavefunctions for spatial states of the antiproton, or neglecting saturation effect of strong forces. The imaginary part of the potential is expected to suppress an antiprotonic wavefunction inside the nucleus and consequently $\Gamma_\mathrm{ST}$ does not grow linearly with the simple overlap ${\cal O}$ as anticipated by \eqref{true_gamma_st}. We further explore the saturation effect by fitting the data to an extended model (described in Appendix) effectively including saturation. This fit is presented Fig.~\ref{Fig1} with a dashed blue line.  Due to the large spread of the experimental data attributed to the neglected contributions, we find that the general linear trend is sufficient for our purposes of a broad overview. However, we motivate future refined work utilizing more sophisticated models or first principle approaches of the antiproton–nucleus interactions.

It should also be noted here that one of the isotopes, $^{58}$Fe, clearly deviates from the general trend and particularly from its neighbors $^{54}$Fe, $^{56}$Fe, and $^{57}$Fe. This may suggest a potential systematic error in the measured decay rate or mislabeled quantum number. No known nuclear structure phenomena could explain this large deviation from the neighboring isotopes.

\section{Stability}

Directly from the assumed model, we can now draw some interesting conclusions about the relevance of different effects depending on the atomic nucleus chosen and the orbit occupied by the antiproton. In Fig.~\ref{Fig2} we collect all these predictions together with experimental points representing different observed antiprotonic atoms~\cite{kanert1986first,trzcinska2001information,bamberger1970observation,poth1978antiprotonic,kreissl1988remeasurement,hartmann2001nucleon,gotta2008x}. For clarity we display only the lowest principal quantum number $n$ achieved in given experiment.

First conclusion follows immediately from the observation that due to a finite size of the nucleus, some expected antiprotonic orbits are located inside the nucleus. In these cases, the model breaks down and no bound state may be formed. By comparing directly the radius of the Bohr orbit $r_n$ with the radius of the nucleus $R(A)$, we find a lower bound for allowed principal quantum numbers of an orbiting antiproton:
\begin{equation} \label{Cond1}
   n > \sqrt{\frac{ZR_0}{a_0}}A^{1/6}\approx 0.2\, Z^{1/2}A^{1/6}.
\end{equation}
This limitation is represented in Fig.~\ref{Fig2} with the black curve. Shaded area below marks parameters for which the antiprotonic orbital radius is within the half-density radius of the nucleus.

However, the prediction of an antiprotonic orbit based on condition \eqref{Cond1} does not assure that the orbit exists. This observation can be understood phenomenologically. Namely, if time $T_{\mathrm{Bohr}}(n)$ needed for a single revolution of the antiproton occupying $n$-th orbit is longer than a typical decay rate from this state then for sure such an orbit cannot be recognized as stable. From the two considered decay processes in our approach only strong interactions effects may lead to this kind of instability. From the model point of view, any additional external electromagnetic interaction has to be weaker than bounding effect of Coulomb interactions, since only then they can be treated as perturbation over stable electromagnetic atomic orbits. To give quantitative arguments we compare the decay rate related to annihilation forced by strong interactions $\Gamma_{\mathrm{ST}}(n)$ with a characteristic {\it ''Bohr's decay rate''} (inverse of Bohr's period)
\begin{equation}
\Gamma_{\mathrm{Bohr}}(n)=\frac{1}{T_{\mathrm{Bohr}}(n)}=\frac{Z^2 \alpha^2}{2\pi}\frac{m_{\bar{p}}c^2}{\hbar}\frac{1}{n^3}.
\end{equation}
It is clear that electromagnetically stable orbits are those which at least fulfill the conditions
\begin{align}
1>\frac{\Gamma_{\mathrm{ST}}(n)}{\Gamma_{\mathrm{Bohr}}(n)}&=\frac{2\pi}{\alpha^2}\frac{\hbar\Gamma_0}{ m_{\bar{p}}c^2}\frac{n^3}{Z^2}{\cal O}(n,A)
.
\end{align}
Since the overlap ${\cal O}(n,A)$ decays much faster with the principal quantum number than $n^{-3}$, for each nucleus ($A$, $Z$) there exist a limiting principal quantum number $n$ for which this ratio is $1$, {\it i.e.}, the Bohr orbit is stable against annihilation in the sense described above. In Fig.~\ref{Fig2} this border is displayed with solid red line. Note that all experimentally observed antiprotonic states are far above this limiting $n$ and in the worst cases the Bohr period is more than an order of magnitude smaller than the annihilation time. Additionally, it should be noted that this border forms an upper bound when saturation effects would be included in the description.

\section{Interplay of interactions}
For the lowest principal quantum numbers $n$ the decay from orbits is dominated by strong interactions. As the antiproton moves further away from the nucleus, the strong interactions rapidly diminish and the dominant role determining the lifetime of the antiproton in the orbit is taken over by electromagnetic interactions --- mainly dipole transitions to lower states. Phenomenologically, the dominant strong interaction is present as long as the following condition is fulfilled
\begin{align}
1>\frac{\Gamma_\mathrm{ST}(n)}{\Gamma_\mathrm{EM}(n)}&=\frac{6}{\alpha^5}\frac{\hbar\Gamma_0}{ m_{\bar{p}}c^2}\frac{1}{Z^4}\left[\frac{n^2(n-1)^2}{2n-1}\right]^3 \frac{{\cal O}(n,A)}{{\cal D}^2(n)}
.
\end{align}
Switching to the transitions dominated by the electromagnetic interaction occurs for those principal quantum numbers for which this ratio is less than $1$. This phenomenological border is marked in Fig.~\ref{Fig2} with a solid blue line. In all experiments with highly excited antiprotonic atoms, purely electromagnetic transitions to lower orbital states are to be expected. In these cases, contributions from nuclear effects are rapidly negligible above the indicated threshold. Consequently, transitions to these states are of particularly good use for probing strong-field QED contributions \cite{baptista2025towards}.

On the other hand, capturing phenomena for which strong forces are manifested requires experimental investigations of cases located in the range between the blue and red lines. In these cases, antiprotonic orbits have still electromagnetic character, but their lifetimes are significantly modified by non-electromagnetic forces.

We note from the spread of experimental data of the lowest observed antiprotonic bound state, that transitions near the stability limits remain unobserved. For instance, measurements of noble gases in Ref. \cite{gotta2008x} were constrained by the limited energy range of the x-ray spectrometer. This reflects the scarcity and limited precision of existing data, emphasizing the need for new refined spectroscopic measurements.

\section{Model Improvements}

While our simplified model provides a valuable systematic overview of antiprotonic atom stability, several refinements could enhance its accuracy, particularly for precise experimental comparisons. These improvements, though not expected to drastically alter the overall trends presented in Fig.~\ref{Fig2}, address key aspects of the electromagnetic and strong interactions.

On the electromagnetic side, incorporating the reduced mass (especially for lighter nuclei) and relativistic corrections (mainly for heavier nuclei) would improve the description of the antiproton's dynamics~\cite{PhysRevA.27.657,PhysRevA.69.042501,PhysRevA.106.042804,PhysRevA.109.022819}. Going beyond our focus on circular orbits, a full treatment of different angular momentum orbits is also desirable since along with decreasing angular momentum they have higher overlap with regions where the nucleus is present~\cite{korobov1997energies,paul2021testing,pachucki2024}.

While considering electromagnetic decays, one can go beyond the dominant E1 transitions considered here and include additional decay channels induced by higher multipoles, Auger transitions~\cite{bacher1988degree}, and in some cases also resonant energy transfer between the antiproton and nucleus~\cite{leon1974e2,batty1978e2,leon1979observation,reidy1985measurements,kanert1986first,1990PhysLettB252.27,1993NuclPhysA561.607, 2004PhysRevC.69.044311,gustafsson2024spin}. They would provide a more complete picture of the de-excitation process.

On the nuclear model side, a more realistic description of distributions of neutrons and protons ~\cite{andrae2000finite,hu2022ab} (accounting also for nuclear shell effects, deformations and polarization~\cite{kohler1986precision, liu2025probing}) would refine the calculation of the crucial overlap integral. In particular, cases of short-lived nuclei with pronounced nuclear periphery are expected to significantly alter the stability~\cite{aumann2022puma}.

Finally, regarding the strong interaction, more sophisticated models, going beyond the zero-range approximation and simplified overlap calculations are ongoing efforts~\cite{batty1981optical,kalbermann1989finite}. This could involve a finite-range potentials~\cite{klos2007neutron} and modeling of bound states formed after annihilation~\cite{schmidt2024production}. These refinements, while not significantly impacting the qualitative global trends presented in Fig.~\ref{Fig2}, are crucial for interpreting specific experimental results and achieving high-precision tests of fundamental physics.

\section{Conclusions}

In this work, we explore the stability of antiprotonic orbits near the nucleus against the influence of electromagnetic and strong interaction decays. By modeling the system with a hydrogen-like framework with a simplified overlap integral for the strong interaction, we have uncovered the dominant factors influencing decay rates across a range of nuclei at a phenomenological level.

By fitting X-ray data of antiprotonic atoms with a two-parameter Fermi density, we obtain an average nuclear diffuseness of $\sigma = 1.02$ fm with an effective strong interaction strength parameter $\hbar\Gamma_0 = (12.1 \pm 0.6)$ MeV. The linear trend approximated in $\Gamma_{\text{ST}}$ reinforces the applicability of simple assumptions for phenomenological estimates. We observed no evident indications of the strong interaction saturation from measurements of the deepest states probed using X-ray spectroscopy. Based on this, we construct a stability diagram that identifies regions where strong interactions or electromagnetic decays dominate, guiding future X-ray spectroscopic studies. While detailed predictions require a sophisticated treatment of the antiproton–nucleus interaction, our map offers a qualitative overview that remains robust to model refinements. The scarcity of data near the predicted stability boundaries highlights the need for new, high-precision measurements, which could provide valuable benchmarks for testing fundamental interactions and refining theoretical models. 

All numerical data presented in this paper are freely available online~\cite{Zenodo}.

\section*{Acknowledgments}
The authors are very grateful to Ben Ohayon for his fruitful comments and suggestions. The authors are also indebted for many valuable discussions with AEgIS Collaboration members. The work by DP and TS was supported as part of a project funded by the Polish Ministry of Education and Science on the basis of Agreement No. 2022/WK/06.

\appendix* 
\section{Role of saturation effects} \label{sec:appendix}
To estimate the impact of strong interaction saturation effects on the stability map (Fig. \ref{Fig2}), neglected in our primary minimal model~\eqref{overlap_approx}, we introduce a phenomenological modification to the strong interaction width. As discussed in  \cite{NPAGalFriedmanSaturation,NPABattyFriedmanGalUnified}, strong absorption leads to an effective repulsion of the antiproton wavefunction from the nucleus. Consequently, it causes the width $\Gamma_\mathrm{ST}$ to saturate rather than increase linearly with the overlap integral ${\cal O}(n,A)$. Therefore, to take this effect into account, we refine the phenomenological relation \eqref{overlap_approx} to the form:
\begin{equation} \label{eq:gamma_sat}
\widetilde\Gamma_{\text{ST}}(n, A) = \frac{\widetilde\Gamma_0 \cdot {\cal O}(n, A)}{1 + \gamma \cdot {\cal O}(n, A)},
\end{equation}
where $\gamma$ is an additional, dimensionless parameter. This generalized form ensures that $\widetilde\Gamma_\mathrm{ST}$ approaches $\widetilde\Gamma_0 \cdot {\cal O}(n,A)$ for small overlaps (negligible saturation) and saturates towards $\widetilde\Gamma_0/\gamma$ for very large overlaps. The parameter $\gamma$ determines how quickly saturation sets in.
Repeating our fitting procedure with this new ansatz we obtain the fitted parameters, $\hbar\widetilde\Gamma_0 = (13.2\pm1.2 )\,\mathrm{MeV}$ and $\gamma=(1.6\pm 1.6)\times 10^3$. With the increased degree of freedom, the fit uncertainties of both $\widetilde\Gamma_\mathrm{0}$ and $\gamma$ are large, highlighting that the data quality is not sufficient to resolve any hint of saturation in the x-ray spectroscopically accessible transitions. 

Finally, to expose the robustness of our map in Fig.~\ref{Fig2}, we use the new values of $\widetilde\Gamma_0$ and $\gamma$ to recalculate the ratios $\widetilde\Gamma_{\text{ST}} / \Gamma_{\text{Bohr}}$ and $\widetilde\Gamma_{\text{ST}} / \Gamma_{\text{EM}}$ and compare with corresponding ratios calculated previously. From this comparison, we find no visible difference between these two approaches and the overall qualitative structure of the map remains robust~\cite{note}. The clear distinction between a region dominated by electromagnetic decays (large $n$, low $Z$) and a region dominated by strong interaction effects (low $n$, high $Z$) persists. The characteristic slopes and relative positions of the contour lines are the same, indicating that the fundamental competition between the $n$-dependence of electromagnetic and strong interaction processes is captured even by the simpler model. With this comparison, we can safely conclude that while the precise quantitative boundaries shown in Fig.~\ref{Fig2} are sensitive to saturation effects (and other simplifications in our model), the global overview of stability conditions provides a useful and qualitatively robust picture of the interplay between electromagnetic and strong decays in antiprotonic atoms, consistent with the first-order aims of our minimal framework.

\bibliography{biblio}

\end{document}